\documentclass[11pt,tightenlines,eqsecnum,floats,aps,amsmath,amssymb,nofootinbib,prd,shownopacs,floatfix]{revtex4}

\usepackage{amsmath,amssymb,amsfonts,dcolumn,color,graphicx,graphics,latexsym}
\usepackage[mathscr]{eucal}
\usepackage[latin1]{inputenc}
\usepackage{enumerate}
\newcommand{\be}{\begin{equation}}
\newcommand{\ee}{\end{equation}}
\newcommand{\ba}{\begin{eqnarray}}
\newcommand{\ea}{\end{eqnarray}}

\newcommand{\bmult}{\nopagebreak[3]\begin{multline}}
\newcommand{\emult}{\end{multline}}

\begin{document}

\title{Generic absence of strong singularities in loop quantum Bianchi-IX spacetimes}
\author{Sahil Saini}
\email{ssaini3@lsu.edu}
\author{Parampreet Singh}
\email{psingh@phys.lsu.edu}
\affiliation{ Department of Physics and Astronomy,\\
Louisiana State University, Baton Rouge, LA 70803, U.S.A.}

\begin{abstract}
We study the generic resolution of strong singularities in loop quantized effective Bianchi-IX spacetime in two different quantizations - the connection operator based `A' quantization and the extrinsic curvature based  `K'  quantization. 
We show that in the effective spacetime description with arbitrary matter content, it is necessary to include inverse triad corrections to resolve all the strong singularities in the `A' quantization. 
Whereas in the `K'  quantization these results can be obtained without including inverse triad corrections. Under these conditions, the energy density, expansion and shear scalars for both of the quantization 
prescriptions are bounded. Notably, both the quantizations can result in potentially curvature divergent events if matter content allows divergences in the partial derivatives of the energy density with respect to the 
triad variables at a finite energy density. Such events are found to be weak curvature singularities beyond which geodesics can be extended in the effective spacetime. Our results show that 
all potential strong curvature singularities of the classical theory are forbidden in Bianchi-IX spacetime in loop quantum cosmology and geodesic evolution never breaks down for such events.  
\end{abstract}

\maketitle

\section{\bf Introduction}

Despite being the most successful theory of classical gravity, which is now empirically known to describe the dynamics of spacetime from the very early universe to the merger of black holes,  Einstein's theory of General Relativity (GR) also predicts the scenarios of its own break down. These are the situations where spacetime curvature components diverge, tidal forces become infinite and geodesic evolution is incomplete.  The singularity theorems  show that such scenarios are unavoidable provided the matter content satisfies certain reasonable energy conditions \cite{hawking-book}. Moreover, cosmological observations point to a universe that would have an initial singularity as per GR. A similar singularity is also believed to be the end state of the gravitational collapse. It is widely expected that a theory of quantum gravity, incorporating the quantum structure of spacetime, will provide a satisfactory resolution to the problem of singularities as the latter are precisely the kind of extreme events where quantum effects are expected to play a crucial role. It is quite possible that the singularities in GR are a result of assuming a continuum differentiable spacetime manifold at all the scales, and a theory of quantum gravity with a quantum structure of space and time may overcome this fundamental problem.  

There are various ways of identifying and classifying the types of singularities that occur in GR. Of these, geodesic incompleteness is considered to be the most general feature of singularities. A more physical characterization of singularities is based on the abnormal behavior of the spacetime curvature near the singular events, on the basis of which singularities are classified into weak and strong singularities. A strong curvature singularity is one where any in-falling object is crushed to zero volume in a finite time regardless of the properties of the object \cite{ellis1977singular,tipler1977singularities,ck-1985conditions}. In contrast, weak curvature singularities may be associated with divergence in one or more components of the Riemann tensor, but tidal forces are not strong enough to crush an arbitrarily strong in-falling object. Further, geodesics can be extended beyond them \cite{clarke-book,Fernandez-Jambrina2006,ps09}. In this sense, weak curvature singularities turn out to be harmless, at least in 
various spacetimes studied so far.

Loop Quantum Gravity (LQG), as a candidate theory of quantum gravity, provides an interesting avenue to test the fate of singularities when quantum gravity effects are included. Investigations in this direction have been carried out in symmetry reduced models in Loop Quantum Cosmology (LQC) \cite{as-status}, and it has been found that the quantum evolution replaces the cosmological singularities by a quantum bounce in various isotropic \cite{aps3, aps1, aps2, slqc, bp-lambda, apsv, closed-warsaw, kv-open, Szulc2007, kp-lambda, madrid-comp, ck-closed1, Martin-Benito2009, ap-lambda, rad}, anisotropic \cite{awe-bianchi1, awe-bianchi2, we-bianchi9, pswe, numlsu-4}, black hole \cite{yks}, and inhomogeneous spacetimes \cite{Martin-Benito2010,Garay2010}. Surprisingly, it has been found that the quantum evolution is well approximated by an effective continuum spacetime with modified Einstein's equations which include quantum gravity effects. This effective description has been numerically tested in various investigations \cite{aps3,ps12,numlsu-4,numlsu-1, numlsu-2, numlsu-3} and shown to extremely closely approximate the quantum evolution including the bounce regime. This continuum effective description thus provides a convenient stage to understand  the effects of quantum gravity on the strength of singularities and geodesic evolution. It has been shown for the spatially flat isotropic and homogeneous models \cite{ps09}, spatially curved models \cite{psfv}, Bianchi-I spacetime \cite{ps11}, Kantowski-Sachs spacetime \cite{ks-strong} and Bianchi-II spacetime \cite{Saini2017} with arbitrary matter that the effective spacetime is geodesically complete and strong singularities do not occur. While the expansion and shear scalars are found to be universally bounded in all cases \cite{cs-geom,ps09,bgps-spatial,ps-proc,ps11,ck-closed2,pswe,ks-bound,cs-schw,ks-strong,cuervo-paper}, the curvature invariants and the time derivative of the expansion scalar may still diverge \cite{wands,ps09,ps11,ks-strong,gowdy-sing}. However, all these divergences have been shown to correspond to weak singularities which are harmless \cite{ps09,psfv,ps11,ks-strong,Saini2017}.

In this manuscript, we aim to extend the results on the generic resolution of strong curvature singularities to the effective description of loop quantized Bianchi-IX spacetime. 
It is well known that Bianchi-IX spacetimes are of special importance in the study of generic approach to singularities due to the conjecture of 
Belinski-Khalatnikov-Lifshitz (BKL) \cite{bkl1971, Berger1998, Garfinkle2004, Garfinkle2004a, Curtis2005, garfinkle, Reiterer2010}. The conjecture states that in the vicinity of a generic 
space-like singularity, the time derivatives dominate over the space derivatives and each spatial point evolves independently of its neighboring points. In the approach to generic space-like singularities, 
the spacetime resembles a Bianchi-IX spacetime. Essentially, as the singularity is approached, the spacetime at each spatial point goes through an infinite sequence of Kasner epochs, where each Kasner epoch is a particular Bianchi-I spacetime, 
and the transitions between these Kasner epochs are well approximated by Bianchi-II spacetimes. These chaotic transitions in the approach to singularity are referred to as the Mixmaster behavior. Previously it has been shown that the 
effective dynamics of Bianchi-I \cite{ks-strong} and Bianchi-II spacetime \cite{Saini2017} are free from strong singularities. To show generic resolution of singularities in general situations in LQC, an important gap was the proof for the Bianchi-IX spacetime. This manuscript aims to fill this gap.
As in previous studies, this work assumes that the effective spacetime description of LQC remains faithful to the quantum evolution. The validity of this assumption has been numerically tested in both isotropic \cite{numlsu-1, numlsu-2, numlsu-3} and anisotropic models \cite{numlsu-4}. 

The loop quantization of Bianchi-IX spacetime comes with certain quantization ambiguities. There are two different available quantizations for this spacetime which are consistent in the sense that they lead to GR at infra-red scale and yield quantum gravitational effects at a well defined ultra-violet scale. The first of these is based on expressing the classical Hamiltonian constraint in term of connection operator, which can be written in terms of holonomies over open loops, and then one quantizes using techniques of loop quantum gravity \cite{we-bianchi9}. This leads to the connection operator based `A' quantization because it uses the operator corresponding to Ashtekar-Barbero connection $ A^i_a $ as a key object. Note that one has  to consider open loops because expressing the field strength in terms of holonomies over closed loops does not yield an algebra of almost periodic functions. The second way to quantize is to use the fact that due to homogeneity of the Bianchi spacetimes, it is possible to treat the extrinsic curvature $ K^i_a $ as a connection for evaluating the holonomies around open loops. The quantization obtained from this procedure is called the `K' quantization. This quantization was first discussed in Refs. \cite{kevin-no-bound,date,kevin-thesis} and later developed in Ref. \cite{pswe}. We will consider both these quantizations in this manuscript as there are important differences in their effective dynamics. Such differences and their impact on generic resolution of singularities has already been noted in the loop quantized Bianchi-II spacetime \cite{Saini2017}.

In contrast to the non-compact spatially flat models, quantization of spatially curved models such as Bianchi-IX spacetimes has another subtlety related to quantization procedure. Let us recall that in loop quantum gravity quantum geometric effects enter through non-local form of curvature encoded via holonomies and inverse scale factor (or triad) modifications. The latter modifications were earlier found to be not important for the case of singularity resolution in isotropic models \cite{apsv}.\footnote{It is however interesting to  note that singularity resolution in spatially curved isotropic models can be obtained just with inverse scale factor modifications \cite{st} with interesting phenomenological ramifications for cosmological perturbations \cite{thermal}.} But, their significance has been noted for spatially curved anisotropic models \cite{bgps-spatial, corichi-bianchi3, Saini2017}. In particular, in the case of the `A' quantization, it has been found that just including holonomy corrections in the effective Hamiltonian constraint is not enough, and one has to further impose weak energy conditions to obtain generic singularity resolution \cite{Saini2017}. This is in contrast to the cases of isotropic, Bianchi-I and Kantowski-Sachs spacetimes where generic  singularity resolution could be obtained just with the holonomy corrections. However, as we will show in Sec. III that even imposition of weak energy conditions on the matter content is not enough to ensure boundedness of expansion and shear scalars in Bianchi-IX spacetimes. If one insists on working with the `A' quantization, then including inverse triad corrections is indispensable for avoiding the singularities. Though the importance of inverse triad corrections for spatially curved models has been noted before, our results clearly establish that the inverse triad corrections are essential for a generic singularity resolution  if one chooses `A' quantization.

In the case of Bianchi-II spacetime, above difficulties with the `A' quantization led to the reconsideration of the `K'  quantization. As discussed above, this quantization results from treating the extrinsic curvature as a connection, which can be done in homogeneous spacetimes by gauge fixing. It was noted in Ref. \cite{pswe} that the effective dynamics of the `K'  quantization leads to generic bounds in expansion and shear scalars without including inverse triad corrections or imposing any energy conditions. Quite expectedly, it was found in Bianchi-II case  that in the `K' quantization, singularity resolution and geodesic completeness was obtained without the need of introducing inverse triad corrections unlike the `A' quantization \cite{Saini2017}. We will find that as in the Bianchi-II case \cite{Saini2017}, the `K' quantization maintains its favorability in the Bianchi-IX case as well and yields generic resolution of singularities without including inverse triad corrections. The triad variables remain finite in  finite time evolution. The directional Hubble rates, expansion and shear scalars also remain finite in finite time evolution. And there are no strong singularities even without including inverse triad modifications. Hence, we find that the `K'  quantization provides a convenient, and probably a superior alternative to the `A' quantization as far as singularity resolution is concerned.

This manuscript is organized as follows. Sec. II gives a brief summary of the classical Hamiltonian description of Bianchi-IX spacetime in terms of Ashtekar variables. This section also provides the expressions for some key quantities such as expansion and shear scalars, the directional Hubble rates and the time derivative of the expansion scalar. Sec. III covers the effective dynamics of the `A' quantization of Bianchi-IX spacetime. This is divided into two parts: Sec. IIIA deals with the `A' quantization without inverse triad corrections, and Sec. IIIB includes the effects of inverse triad corrections in the effective dynamics. We show in Sec. IIIA, that vanishing of triads in finite time evolution can not be avoided if inverse triad corrections are not included, and that imposing weak energy condition, in addition, on the energy density is not sufficient to prevent such a phenomena either. Sec. IIIB shows that  in the `A' quantization when inverse triad corrections are included, none of the triads vanish or diverge in a finite time. In Sec. IV, we describe the effective dynamics of the `K'  quantization for Bianchi-IX model, and we find that the energy density, and the expansion and shear scalars are bounded without requiring any energy condition or inverse triad modifications. The triads never vanish or diverge in finite time evolution without requiring inverse triad modifications in the effective Hamiltonian. However, we find in Sec. V, that the curvature invariants, components of Riemann tensor and time derivative of the expansion scalar can still potentially diverge both in `A' and `K' quantizations. This does not imply existence of singularities. We show in Sec. VI that for the `A' quantization with inverse triad correction and the `K'  quantization even without the inverse triad modifications, the geodesics are well behaved when such divergences in curvature invariants occur. In Sec. VII, we show that in finite time evolution no strong singularities exist in the effective dynamics of the `K'  quantization or the `A' quantization with inverse triad corrections. The events where curvature invariants may potentially diverge thus only amount to weak singularities. In Sec. VIII we provide a summary of our results.

\section{Classical aspects of Bianchi-IX spacetime in connection-triad variables}
The Bianchi models are a class of anisotropic cosmological spacetimes where the spatial hypersurfaces are homogeneous. In case of Bianchi-IX model, the hypersurfaces have a topology of $\mathcal{S}^3$ along with the symmetries associated with the three rotations within the hypersurface. The metric for the Bianchi-IX model can simply be written as \cite{pswe}:
\be
\mathrm{d} s^2= - \mathrm{d} t^2 + q_{a b} \mathrm{d} x^a \mathrm{d} x^b ~.
\ee
The physical metric on spatial manifold can be written as $q_{ab}=\omega^i_a \omega^j_b \delta_{ij}$, where the physical forms are $\omega ^i_a=a^i (t) ^o\omega^i_a$, and the fiducial ones are given by
\begin{eqnarray}
^o\omega^1_a &=& \sin \beta \sin \gamma (\mathrm{d}\alpha)_a + \cos \gamma (\mathrm{d}\beta)_a, \\
^o\omega^2_a &=& -\sin \beta \cos \gamma (\mathrm{d}\alpha)_a + \sin \gamma (\mathrm{d}\beta)_a, \\
^o\omega^3_a &=& \cos \beta (\mathrm{d}\alpha)_a + (\mathrm{d}\gamma)_a.
\end{eqnarray}
where $\alpha$, $\beta$ and $\gamma$ are the angular coordinates on a 3-sphere with a radius $r_o= 2$ and fiducial volume $V_0 = l_o^3 = 2 \pi^2 r_o^3$. 

The phase space variables in loop quantum gravity are the Ashtekar-Barbero connection $A_{a}^{i}$ and the conjugate triad $E_{i}^{a}$. After symmetry reduction, these can be given in terms of the fiducial triads ${\mathring e}_{i}^{a}$ and co-triads ${\mathring\omega}_{a}^{i}$ as:
\begin{equation}
E_{i}^{a}=\frac{p_i}{l_o}\sqrt{|\mathring q|} {\mathring e}_{i}^{a} \quad \text{and} \quad A_{a}^{i}=\frac{{c}^i}{l_o} {\mathring\omega}_{a}^{i}
\end{equation}
where $\mathring q$ is the determinant of the spatial metric. The non-zero Poisson brackets between the symmetry reduced connection $c^i$ and triads $p_i$ and are given by
\begin{equation}
\left\lbrace c^i,p_j \right\rbrace =8\pi G \gamma \delta_{j}^{i} .\label{poisson1}
\end{equation}
Here, the triad variables are related to the directional scale factors through relations of the type
\begin{equation}
p_1 = \mathrm{sgn}(a_1)|a_2 a_3| {l_o}^2, \label{scale factors to triads}
\end{equation}
(and similarly for $p_2$ and $p_3$). The relationship between the connection components and the time derivative of the metric components is found via the Hamilton's equations using the classical Hamiltonian. For the choice of lapse $N=1$, it can be written as \cite{we-bianchi9},
\begin{eqnarray}
\mathcal{H}_{\mathrm{cl}} &=& -\frac{1}{8 \pi G\gamma^2 \sqrt{|p_1 p_2 p_3|}}\bigg[p_1 p_2 c_1 c_2 + p_2 p_3 c_2 c_3 + p_3 p_1 c_3 c_1 + l_o \epsilon (p_1 p_2 c_3 + p_2 p_3 c_1 + p_3 p_1 c_2 )\nonumber \\
&& +\frac{l_o^2 (1+\gamma^2)}{4}\bigg(2 p_1^2+2p_2^2+2p_3^2-\frac{p_1^2 p_2^2}{p_3^2}-\frac{p_2^2 p_3^2}{p_1^2}-\frac{p_3^2 p_1^2}{p_2^2} \bigg)\bigg]+\rho\sqrt{|p_1 p_2 p_3|}, \label{C_Hamiltonian}
\end{eqnarray}
where $\rho$ is the energy density of a minimally coupled matter field, and $\gamma \approx 0.2375$ is the Barbero-Immirzi parameter whose value is fixed by black hole thermodynamics in loop quantum gravity. Here the constant $\epsilon = \mathrm{sgn}(p_1) \mathrm{sgn}(p_2) \mathrm{sgn}(p_3)$ is either $+1$ or $-1$ depending on whether the triads are right-handed or left-handed respectively. Note that reflections in the triad space leave the above Hamiltonian invariant because each term is invariant under change in the orientation in any of the physical triads. 

The energy density on the constraint surface is given by the vanishing of the Hamiltonian constraint ${\cal C}_{\mathrm{cl}} = 8 \pi G {\cal H}_{\mathrm{cl}} \approx 0$. Since the energy density only depends on the magnitudes of the triads for the same reason as above, we can write down its expression in the positive octant:
\begin{eqnarray}
\rho &=& \frac{1}{8 \pi G\gamma^2 p_1 p_2 p_3}\bigg[p_1 p_2 c_1 c_2 + p_2 p_3 c_2 c_3 + p_3 p_1 c_3 c_1 + l_o (p_1 p_2 c_3 + p_2 p_3 c_1 + p_3 p_1 c_2 )\nonumber \\
&& +\frac{l_o^2 (1+\gamma^2)}{4}\bigg(2 p_1^2+2p_2^2+2p_3^2-\frac{p_1^2 p_2^2}{p_3^2}-\frac{p_2^2 p_3^2}{p_1^2}-\frac{p_3^2 p_1^2}{p_2^2} \bigg)\bigg].
\end{eqnarray}

Using the relationship between the triads and directional scale factors, it is straightforward to express the 
 directional Hubble rates, the expansion scalar and its time derivative, and the shear scalars  as follows:
\begin{eqnarray}
H_i &=& \frac{\dot a_i}{a_i} = \frac{1}{2} \left(\frac{\dot p_j}{p_j} +\frac{\dot p_k}{p_k} - \frac{\dot p_i}{p_i} \right) \quad \text{where} \quad i,j,k \in \lbrace 1,2,3 \rbrace; i\neq j \neq k, \label{hubblerates} \\
\theta &=& H_1 + H_2 + H_3, \label{expansionscalar} \\
\dot \theta &=& \frac{1}{2} \sum_{i=1}^{3} \bigg(\frac{\ddot p_i}{p_i} - \bigg(\frac{\dot p_i}{p_i} \bigg)^2 \bigg), \label{thetadot} \label{expansionrate} ~~~~ \mathrm{and} ~~~~\\
\sigma ^2 &=& \frac{1}{3} \bigg( (H_1-H_2)^2 + (H_2-H_3)^2 + (H_3-H_1)^2 \bigg). \label{shearscalar}
\end{eqnarray}

Similarly the Ricci scalar, Kretschmann scalar and other curvature invariants can be obtained in terms of the triad variables and their time derivatives. The equations of motion for the triad variables thus determine the time evolution for all the quantities of interest in the analysis of singularities.

We can use the classical Hamiltonian \eqref{C_Hamiltonian} to obtain the equations of motion for the triads, which in turn give the time evolution for all the other quantities of interest:
\begin{eqnarray}
\dot p_1 &=& \frac{p_1}{\gamma \sqrt{|p_1 p_2 p_3|}} \bigg(p_2 c_2 + p_3 c_3 + l_o \epsilon \frac{p_2 p_3}{p_1} \bigg), \\
\dot p_2 &=& \frac{p_2}{\gamma \sqrt{|p_1 p_2 p_3|}} \bigg(p_3 c_3 + p_1 c_1 + l_o \epsilon \frac{p_3 p_1}{p_2} \bigg), \\
\dot p_3 &=& \frac{p_3}{\gamma \sqrt{|p_1 p_2 p_3|}} \bigg(p_1 c_1 + p_2 c_2 + l_o \epsilon \frac{p_1 p_2}{p_3} \bigg).
\end{eqnarray}
These classical equations of motion are known to lead to singularities in the classical Bianchi-IX spacetime. Specifically, the time evolution of the triad variables causes some of them to either vanish or diverge, causing divergences in the curvature invariants, energy density, Hubble rates, expansion and shear scalars. As an example, starting from matter obeying weak energy condition the Bianchi-IX spacetime undergoes an evolution such that the anisotropic shear dominates, unless the matter is a massless scalar or with an ultra-stiff equation of state.
The spacetime undergoes a Mixmaster dynamics before reaching a cigar type singularity. The situation for the singularity is similar even if the initial conditions are such that the approach to the singularity is isotropic which can happen in the case of massless scalar and ultra-stiff matter. A consequence of vanishing of the triads in finite time evolution is that even for reasonable matter-energy conditions, the spacetime curvature diverges in finite time and geodesic evolution stops. We will see in the next two sections, the way the equations of motion for triad variables are modified in effective dynamics of LQC which ensures that the time evolution of triad variables does not lead them to vanish or diverge. This result has profound consequences for generic resolution of singularities in LQC.

\section{Effective loop quantum cosmological dynamics: ~~~~~~~~~~~~~~~~~~~~~~~~~~`A' Quantization}

In this section, we will explore the effective dynamics of the Bianchi-IX spacetime in the connection operator based `A' quantization, first studied in \cite{we-bianchi9}. Let us recall that the quantum corrections enter the effective Hamiltonian in LQC in two different forms - holonomy corrections and inverse triad corrections. In the previous studies of effective dynamics of cosmological models, it was observed that a generic resolution of strong curvature singularities could be obtained by just considering the holonomy corrections in the case of isotropic and homogeneous models \cite{ps09,psfv}, as well as in Bianchi-I \cite{ps11} and Kantowski-Sachs models \cite{ks-strong}. However in Bianchi-II models \cite{Saini2017}, it was seen that inverse triad corrections are essential for proving singularity resolution in the `A' quantization, otherwise imposing weak energy conditions on the energy density was required if inverse triad corrections were to be ignored. In the case of effective dynamics of Bianchi-IX model in the `A' quantization, we will see that even imposing weak energy conditions is not enough to get the desired results if inverse triad corrections are ignored \cite{bgps-spatial}. We will obtain the dynamics without inverse triad corrections in Sec. IIIA below and point out the possible scenarios leading to singularities that are not necessarily avoided in this case. These scenarios result from vanishing or divergence of at least one of the triads in finite time. Then in Sec. IIIB,  we show that if inverse triad corrections are included, then triads remain non-zero and finite in a finite time evolution, signaling resolution of singularities. This indicates that if we work with `A' quantization, then inverse triad corrections are absolutely essential for singularity resolution in a general cosmological model with anisotropies. However, as we show in the next section, there is an alternative quantization called the `K'  quantization in which these results are easily obtained for Bianchi-IX model without needing to include inverse triad corrections or imposing any energy conditions on the energy density.

In our analysis in this section, we will restrict ourselves to the positive octant in the triad space for convenience after introducing the equations of motion. In a quantization similar to \cite{awe-bianchi2}, choosing those quantum states which are reflection symmetric in triad space allows one to restrict to the positive octant without any loss of generality. It is also possible to choose a different factor ordering in the Hamiltonian constraint, as is done in \cite{Martin-Benito2010, Garay2010} for the Bianchi-I case, such that different quadrants decouple and one can safely choose to work in one of them. We point this out again in the following subsection when we restrict to the positive octant. The effective Hamiltonian in the positive octant will be the same arising from either of the above mentioned approaches. 

\subsection{`A' quantization without inverse triad corrections}

We start with the expression for the effective Hamiltonian following the construction in \cite{we-bianchi9}, where we have kept the signs of all the terms explicit in the expression. The effective Hamiltonian for $N=1$ with minimally coupled matter can be written as follows:
 \begin{eqnarray}
\mathcal{H} &=& -\frac{\sqrt{|p_1 p_2 p_3|}}{8 \pi G\gamma^2 \lambda^2}\bigg[(\mathrm{sgn}(p_1)\mathrm{sgn}(p_2)\sin(\bar\mu_1 c_1)\sin(\bar\mu_2 c_2)+ \mathrm{cyclic}) + l_o \lambda \bigg( \mathrm{sgn}(p_3)\sqrt{\frac{|p_1 p_2|}{{|p_3|}^3}}\sin(\bar\mu_3 c_3) \nonumber \\ 
&& + \mathrm{cyclic}\bigg) +\frac{l_o^2 \lambda^2 (1+\gamma^2)}{4 |p_1 p_2 p_3|}\bigg(2 p_1^2+2p_2^2+2p_3^2-\frac{p_1^2 p_2^2}{p_3^2}-\frac{p_2^2 p_3^2}{p_1^2}-\frac{p_3^2 p_1^2}{p_2^2} \bigg)\bigg]+\rho\sqrt{|p_1 p_2 p_3|}. \label{Hamiltonian_A}
\end{eqnarray}
The equations of motion for triads are:
\begin{eqnarray}
\dot p_1 &=& \frac{p_1}{\gamma \lambda} \bigg( \mathrm{sgn}(p_2) \sin(\bar\mu_2 c_2) + \mathrm{sgn}(p_3) \sin(\bar\mu_3 c_3) + l_o \lambda \frac{\sqrt{|p_2 p_3|}}{|p_1|^{3/2}}\bigg) \cos(\bar\mu_1 c_1), \label{p_1_A} \\
\dot p_2 &=& \frac{p_2}{\gamma \lambda} \bigg( \mathrm{sgn}(p_3) \sin(\bar\mu_3 c_3) + \mathrm{sgn}(p_1)\sin(\bar\mu_1 c_1) + l_o \lambda \frac{\sqrt{|p_3 p_1|}}{|p_2|^{3/2}}\bigg) \cos(\bar\mu_2 c_2), \label{p_2_A} \\
\dot p_3 &=& \frac{p_3}{\gamma \lambda} \bigg(\mathrm{sgn}(p_1) \sin(\bar\mu_1 c_1) + \mathrm{sgn}(p_2)\sin(\bar\mu_2 c_2) + l_o \lambda \frac{\sqrt{|p_1 p_2|}}{|p_3|^{3/2}}\bigg) \cos(\bar\mu_3 c_3). \label{p_3_A}
\end{eqnarray}
Here $\lambda^2= \Delta = 4 \sqrt{3} \pi \gamma l_{\mathrm{pl}}^2$ denotes minimum area eigenvalue from the underlying quantum geometry, and
\be
\bar\mu_1=\lambda \sqrt{\frac{|p_1|}{|p_2 p_3|}}, \quad \bar\mu_2=\lambda \sqrt{\frac{|p_2|}{|p_1 p_3|}}, \quad \bar\mu_3=\lambda \sqrt{\frac{|p_3|}{|p_1 p_2|}}. \label{mubar}
\ee

\noindent
Let us consider the eq. \eqref{p_1_A}, which we can formally integrate as:
\begin{equation}
\int_{p_1^0}^{p_1(t)}\frac{\mathrm{d}p_1}{p_1}= \int_{t_0}^t \frac{1}{\gamma \lambda} \bigg( \mathrm{sgn}(p_2) \sin(\bar\mu_2 c_2) + \mathrm{sgn}(p_3) \sin(\bar\mu_3 c_3) + l_o \lambda \frac{\sqrt{|p_2 p_3|}}{|p_1|^{3/2}}\bigg) \cos(\bar\mu_1 c_1) \mathrm{d}t ~.
\end{equation}
It yields,
\begin{equation}
p_1(t)=p_1^0 \exp\left\lbrace\int_{t_0}^t \frac{1}{\gamma \lambda} \bigg( \mathrm{sgn}(p_2) \sin(\bar\mu_2 c_2) + \mathrm{sgn}(p_3) \sin(\bar\mu_3 c_3) + l_o \lambda \frac{\sqrt{|p_2 p_3|}}{|p_1|^{3/2}}\bigg) \cos(\bar\mu_1 c_1) \mathrm{d}t \right\rbrace .
\end{equation}

Similarly, we can obtain the expressions for $ p_2 $ and $ p_3 $ as:
\begin{eqnarray}
p_2(t) &=&p_2^0 \exp\left\lbrace\int_{t_0}^t \frac{\mathrm{d}t}{\gamma \lambda} \bigg( \mathrm{sgn}(p_3) \sin(\bar\mu_3 c_3) + \mathrm{sgn}(p_1)\sin(\bar\mu_1 c_1) + l_o \lambda \frac{\sqrt{|p_3 p_1|}}{|p_2|^{3/2}}\bigg) \cos(\bar\mu_2 c_2) \right\rbrace , \\
p_3(t) &=& p_3^0 \exp\left\lbrace\int_{t_0}^t \frac{\mathrm{d}t}{\gamma \lambda} \bigg( \mathrm{sgn}(p_1) \sin(\bar\mu_1 c_1) + \mathrm{sgn}(p_2) \sin(\bar\mu_2 c_2) + l_o \lambda \frac{\sqrt{|p_1 p_2|}}{|p_3|^{3/2}}\bigg) \cos(\bar\mu_3 c_3) \right\rbrace .
\end{eqnarray}
It can be seen that singularities are not necessarily avoided by these dynamical equations. For example, the above equations seem to allow a situation when one of the triads, say $p_1$, approaches zero while $p_2$ and $p_3$ remain finite. Or, the scenario where all three of the triads diverge simultaneously. Analytically it is not obvious that above scenarios for singularities can be excluded dynamically by the equations of motion.

The expressions for the Hubble rates, the expansion and shear scalars, and the time derivative of the expansion scalar remain the same in the effective dynamics as in the classical case. They are given by eqs. \eqref{hubblerates}, \eqref{expansionscalar}, \eqref{shearscalar} and \eqref{expansionrate} respectively. We note that all these quantities can diverge in the `A' quantization without inverse triad corrections as one or more of the triad variables can vanish in a finite time evolution.\\

In the following analysis, it is convenient to restrict to the positive octant. The first way to work with the positive octant is to take the approach in Refs. \cite{awe-bianchi1} and \cite{awe-bianchi2} where one chooses the basis states of the kinematical Hilbert space to be reflection-symmetric in the triad space. Since the states are symmetric at the quantum level, we can thus restrict ourselves to the positive octant while obtaining the effective Hamiltonian \cite{awe-bianchi1,awe-bianchi2}. Moreover, we can choose our initial conditions such that the triads start out in the positive octant in our analysis at the effective level. The effective equations of motion for the triads in this section are such that the triads will remain in the same octant where they started out initially until a singularity is reached and one of the triads vanishes, where the equations of motion break down.

Alternatively, the restriction to an octant can be justified merely on the basis of factor ordering while quantizing the Hamiltonian constraint, without requiring the quantum states to be symmetric. In this regard, one can take the path followed in Ref. \cite{Martin-Benito2010,Garay2010} where a different factor ordering of the quantum Hamiltonian constraint is employed so that the different octants are decoupled. This allows us to restrict the triads in the positive octant in our analysis. However, in this approach, the state corresponding to vanishing volume is also decoupled from all the octants. Here let us note that we found from the behavior of triads at the effective level,  that it is possible for one or more triads to vanish in a finite-time evolution starting from the positive octant. This behavior of effective dynamics shows the latter's limitation. The effective Hamiltonian is indifferent to different factor orderings at the quantum level and hence does not know about the effect of decoupling of zero volume state. This limitation is tied to derivation of the effective Hamiltonian which assumes that volumes are greater than the Planck volume \cite{vt}. 

In the rest of this manuscript, we assume that one of the above two choices is applied so that we can focus ourselves on the positive octant in the triad space. Further, let us consider that the initial values of the triads $ p_i^o $ are positive definite. Then the expression of the energy density, from the vanishing of the effective Hamiltonian constraint, is dynamically equal to:
\begin{eqnarray}
\rho &=& \frac{1}{8 \pi G\gamma^2 \lambda^2} \bigg[\sin(\bar\mu_1 c_1)\sin(\bar\mu_2 c_2)+ \mathrm{cyclic}) + l_o \lambda \bigg( \sqrt{\frac{p_1 p_2}{{p_3}^3}}\sin(\bar\mu_3 c_3) + \mathrm{cyclic ~ terms}\bigg) \nonumber \\ 
&&  +\frac{l_o^2 \lambda^2 (1+\gamma^2)}{4 p_1 p_2 p_3}\bigg(2 p_1^2+2p_2^2+2p_3^2-\frac{p_1^2 p_2^2}{p_3^2}-\frac{p_2^2 p_3^2}{p_1^2}-\frac{p_3^2 p_1^2}{p_2^2} \bigg)\bigg].
\end{eqnarray}
It has been shown in Ref. \cite{bgps-spatial} that the above energy density is not bounded above. Moreover, unlike Ref. \cite{Saini2017}, imposing the weak energy condition on the energy density does not prevent any of the triads from vanishing in a finite time evolution. Thus, in contrast to the Bianchi-II case, there is no way the above energy density can be bounded in a generic evolution.

The expression for the Hubble rates can be evaluated using equations \eqref{hubblerates}. The directional Hubble rate $ H_1 $ is given by,
\begin{eqnarray}
H_1 &=& \frac{1}{2 \gamma \lambda} (\sin(\bar\mu_1 c_1 -\bar\mu_2 c_2) + \sin(\bar\mu_1 c_1 -\bar\mu_3 c_3)+\sin(\bar\mu_2 c_2 + \bar\mu_3 c_3) ) \nonumber \\
&& + \frac{l_o}{2\gamma} \bigg(\frac{\sqrt{p_1 p_2}}{p_3^{3/2}} \cos(\bar\mu_3 c_3) + \frac{\sqrt{p_3 p_1}}{p_2^{3/2}}\cos(\bar\mu_2 c_2) - \frac{\sqrt{p_2 p_3}}{p_1^{3/2}}\cos(\bar\mu_1 c_1) \bigg) ~.
\end{eqnarray}
Similarly, the directional Hubble rates $ H_2 $ and $ H_3 $ can be obtained by cyclic permutations of the above equation. We note that the directional Hubble rates can diverge if any of the triad variables diverges or vanishes. This, through equations \eqref{expansionscalar} and \eqref{shearscalar}, produces divergences in expansion and shear scalars as well.

In a similar way, the expressions for the curvature scalars such as the Ricci scalar, the Kretschmann scalar and the square of the Weyl tensor which depend on time derivative so triads can be shown to diverge as one or more triads vanish or diverge. Hence, we find that the `A' quantization without inverse triad corrections is inadequate for the resolution of singularities. Unlike the case of `A' quantization in Bianchi-II spacetime \cite{Saini2017}, even imposing weak energy conditions is not sufficient to prevent the singularities. This implies that it is essential to at least include inverse triad corrections if one hopes to resolve the classical singularities. Indeed, as we will see in the next subsection that including inverse triad corrections helps us obtain the crucial result that none of the triads vanish or diverge in a finite time evolution. This result is crucial to prove generic resolution of singularities.

\subsection{`A' quantization with inverse triad corrections}

The effective Hamiltonian \eqref{Hamiltonian_A} in previous subsection includes the holonomy corrections but ignores the inverse-triad corrections. The inverse triad corrections in the case of lapse $N=V$ have been obtained in \cite{bgps-spatial}. We use the same procedure as in Ref. \cite{bgps-spatial}, to obtain the inverse triad corrections for the case of lapse $N=1$. Let us note that the eigenvalues of the inverse triad operator $\widehat{{\vert p_1 \vert}^{-1/4}}$ are given as follows:
\begin{equation}
\widehat{{\vert p_1 \vert}^{-1/4}} \vert p_1 p_2 p_3 \rangle = g_1(p_1) \vert p_1 p_2 p_3 \rangle,
\end{equation}
where
\begin{equation}
g_1(p_1)=\frac{(p_2 p_3)^{1/4}}{\sqrt{2\pi \gamma \lambda l_{\mathrm{pl}}^2}} ( \sqrt{\vert v+1 \vert} - \sqrt{\vert v-1 \vert}); \quad \quad v=\frac{\sqrt{p_1 p_2 p_3}}{2\pi \gamma \lambda l_{\mathrm{pl}}^2} \label{g} ~.
\end{equation}
Similarly, we can obtain inverse triad corrections for the inverse powers of $p_2$ and $p_3$ which will involve the corresponding functions $g_2$ and $g_3$. Using these expressions, the effective Hamiltonian with inverse triad corrections 
for the $N=1$ case is obtained as follows:
\begin{eqnarray}
\mathcal{H} &=& -\frac{\sqrt{p_1 p_2 p_3}}{8 \pi G\gamma^2 \lambda^2}\bigg[(\sin(\bar\mu_1 c_1)\sin(\bar\mu_2 c_2)+ \mathrm{cyclic}) + l_o \lambda \bigg( \sqrt{\frac{p_1 p_2}{p_3}} g_3^2 \sin(\bar\mu_3 c_3) + \mathrm{cyclic}\bigg)\nonumber \\
&& +\frac{l_o^2 \lambda^2 (1+\gamma^2) (g_1 g_2 g_3)^2}{4 \sqrt{p_1 p_2 p_3}}\bigg(2 p_1^2-p_1^2 p_2^2 g_3^8+ \mathrm{cyclic} \bigg)\bigg]+\rho\sqrt{p_1 p_2 p_3}. \label{Hamiltonian_A_triads}
\end{eqnarray}
The resulting equations of motion for triads are:
\begin{eqnarray}
\dot p_1 &=& \frac{p_1}{\gamma \lambda} \bigg( \sin(\bar\mu_2 c_2) + \sin(\bar\mu_3 c_3) + l_o \lambda \sqrt{\frac{p_2 p_3}{p_1}}g_1^4\bigg) \cos(\bar\mu_1 c_1), \label{p1-triad} \\
\dot p_2 &=& \frac{p_2}{\gamma \lambda} \bigg( \sin(\bar\mu_3 c_3) + \sin(\bar\mu_1 c_1) + l_o \lambda \sqrt{\frac{p_3 p_1}{p_2}}g_2^4\bigg) \cos(\bar\mu_2 c_2), \label{p2-triad} ~~~~ \mathrm{and,} ~~~~\\
\dot p_3 &=& \frac{p_3}{\gamma \lambda} \bigg( \sin(\bar\mu_1 c_1) + \sin(\bar\mu_2 c_2) + l_o \lambda \sqrt{\frac{p_1 p_2}{p_3}}g_3^4\bigg) \cos(\bar\mu_3 c_3). \label{p3-triad} ~.
\end{eqnarray}
We can formally integrate these equations to understand the form of the solution. For example, let $p_1$ starts with some positive-definite value $p_1^o$ at some initial time $t_o$. Then from \eqref{p1-triad} we can write,
\begin{eqnarray}
p_1(t) &=& p_1^0  \exp\left\lbrace\frac{1}{\gamma \lambda} \int_{t_0}^t (\sin(\bar\mu_2 c_2)+\sin(\bar\mu_3 c_3))\cos(\bar\mu_1 c_1) \mathrm{d}t\right\rbrace \nonumber \\
&& \times\exp\left\lbrace \int_{t_0}^t\bigg( \frac{l_o}{\gamma} \sqrt{\frac{p_2 p_3}{p_1}}g_1^4\bigg) \cos(\bar\mu_1 c_1) \mathrm{d}t\right\rbrace. \label{p1-triad-integrated}
\end{eqnarray}
The corresponding equations for $p_2$ and $p_3$ are obtained by cyclic permutations. The directional scale factors are related kinematically to the triad variables through equations of the form \eqref{scale factors to triads}. We are interested to see whether the time evolution of the triads leads them to either vanish or diverge in a finite time evolution. In Table \ref{table:1} we list all the different possibilities for the evolution of the triad variables for all finite values of time $t$ and label the different possibilities from A1 to A10.

\begin{table}[h!]
\centering
	\begin{tabular}{||c | c | c | c | c||}
	\hline
	 & Finite, non-zero & Vanishing & Diverging  & Remarks \\[0.5ex]
	\hline\hline
	A1 & 3 & 0 & 0 & all the triads remain finite and non-zero \\
	\hline
	A2 & 0 & 3 & 0 & all the triads vanish \\
	\hline
	A3 & 0 & 0 & 3 & all the triads diverge \\
	\hline
	A4 & 1 & 1 & 1 & one triad vanishes while one of them diverges \\
	\hline
	A5 & 2 & 1 & 0 & two triads are finite and non-zero while the third vanishes  \\ 
	\hline
	A6 & 2 & 0 & 1 & two triads are  finite and non-zero while the third diverges \\
	\hline
	A7 & 1 & 2 & 0 & two triads vanish while the third remains finite and non-zero \\
	\hline
	A8 & 0 & 2 & 1 & two triads vanish while the third diverges \\
	\hline
	A9 & 1 & 0 & 2 & two triads diverge while the third remains finite and non-zero \\
	\hline
	A10 & 0 & 1 & 2 & two triads diverge while the third vanishes \\ [1ex]
	\hline
	\end{tabular}
\caption{Different possible fates for the triad variables upon evolution}
\label{table:1}
\end{table}

We will show now in the rest of this subsection that the properties of the functions $g_1, g_2$ and $g_3$ defined through \eqref{g} are such that the triads $p_i$ remain non-zero, positive and finite for all finite time evolution. 
That is, except the possibility A1, no other case is possible. We will break down the analysis into three different limits: \\

(i) {\it{The limit of vanishing volume}}: In this case,  $v$ tends to zero and the case A1, A3, A6 and A9 are automatically ruled out. We now show that no other cases are possible either. 
In this limit, it can be seen from \eqref{g} that the functions $g_i$ take the form:
\begin{eqnarray}
g_1 &=& \sqrt{\beta} (p_2 p_3)^{1/4} \bigg(v - \frac{v^3}{8} + ...\bigg), \\
g_2 &=& \sqrt{\beta} (p_3 p_1)^{1/4} \bigg(v - \frac{v^3}{8} + ...\bigg), \\
g_3 &=& \sqrt{\beta} (p_1 p_2)^{1/4} \bigg(v - \frac{v^3}{8} + ...\bigg). \\
\end{eqnarray}
where we have defined for convenience, $\beta = 1/(2\pi \gamma \lambda l_{\mathrm{pl}}^2)$. Since $v$ is vanishing, the terms in the parentheses in RHS of the above equations are all tending to zero. Then equations of motion for triads become,
\begin{eqnarray}
p_1(t) &=& p_1^0  \exp\left\lbrace\frac{1}{\gamma \lambda} \int_{t_0}^t (\sin(\bar\mu_2 c_2)+\sin(\bar\mu_3 c_3))\cos(\bar\mu_1 c_1) \mathrm{d}t\right\rbrace \nonumber \\
&& \times\exp\left\lbrace \frac{l_o \beta^6}{\gamma} \int_{t_0}^t {p_1}^{3/2} (p_2 p_3)^{7/2} \bigg( 1 - \frac{v^2}{8} + ...\bigg)^4 \cos(\bar\mu_1 c_1) \mathrm{d}t\right\rbrace , \label{p1-triad-integrated-1} \\
p_2(t) &=& p_2^0  \exp\left\lbrace\frac{1}{\gamma \lambda} \int_{t_0}^t (\sin(\bar\mu_3 c_3)+\sin(\bar\mu_1 c_1))\cos(\bar\mu_2 c_2) \mathrm{d}t\right\rbrace \nonumber \\
&& \times\exp\left\lbrace \frac{l_o \beta^6}{\gamma} \int_{t_0}^t {p_2}^{3/2} (p_3 p_1)^{7/2} \bigg( 1 - \frac{v^2}{8} + ...\bigg)^4 \cos(\bar\mu_2 c_2) \mathrm{d}t\right\rbrace , \label{p2-triad-integrated-1}, ~~~ \mathrm{and} ~~~ \\
p_3(t) &=& p_3^0  \exp\left\lbrace\frac{1}{\gamma \lambda} \int_{t_0}^t (\sin(\bar\mu_1 c_1)+\sin(\bar\mu_2 c_2))\cos(\bar\mu_3 c_3) \mathrm{d}t\right\rbrace \nonumber \\
&& \times\exp\left\lbrace \frac{l_o \beta^6}{\gamma} \int_{t_0}^t {p_3}^{3/2} (p_1 p_2)^{7/2} \bigg( 1 - \frac{v^2}{8} + ...\bigg)^4 \cos(\bar\mu_3 c_3) \mathrm{d}t\right\rbrace . \label{p3-triad-integrated-1}
\end{eqnarray}
Let us note that in eq. \eqref{p1-triad-integrated-1}, the terms in the first line on RHS are bounded for all finite time because the integrand in the exponential is a bounded function. 
In the limit of vanishing $v$, all the terms in the integrand in the second exponential are bounded except for the factor ${p_1}^{3/2} (p_2 p_3)^{7/2}$. 
We thus find that in order for $p_1$ to either vanish or diverge in a finite time, the factor ${p_1}^{3/2} (p_2 p_3)^{7/2}$ in the exponential must diverge. 
Similarly for $p_2$ and $p_3$ to either vanish or diverge, the corresponding factors in their expressions, given respectively by   ${p_2}^{3/2} (p_3 p_1)^{7/2}$ and  ${p_3}^{3/2} (p_1 p_2)^{7/2}$ must diverge. 
Using this we can easily see that none of the remaining  possibilities from Table \ref{table:1} are consistent with the equations of motion of triads in the case of vanishing volume. 
For example, the case A2 corresponds to all vanishing triads. But from the argument in the above paragraph, for $p_1$ to vanish in finite 
time, ${p_1}^{3/2} (p_2 p_3)^{7/2}$ must diverge in finite time. Similarly, for $p_2$ and $p_3$ to vanish, ${p_2}^{3/2} (p_3 p_1)^{7/2}$ and  ${p_3}^{3/2} (p_1 p_2)^{7/2}$ must diverge 
respectively. We thus reach an inconsistency, and the case A2 is not physically possible. Similarly, A4, A5, A7 and A8 choices in Table I are also ruled out. Thus, none of the cases in Table I allow a vanishing volume consistent with the equations of motion.\\

(ii) {\it{The diverging volume limit:}} In this case,  $v$ tends to infinity and the cases A1, A2, A5 and A7 in Table I are automatically ruled out. In this limit, the expressions for $g_i$ become:
\begin{eqnarray}
g_1 &=& \frac{1}{p_1^{1/4}} \bigg( 1 - \frac{1}{8 v^2} + ... \bigg), \\
g_2 &=& \frac{1}{p_2^{1/4}} \bigg( 1 - \frac{1}{8 v^2} + ... \bigg), ~~~~\mathrm{and} ~~~~\\
g_3 &=& \frac{1}{p_3^{1/4}} \bigg( 1 - \frac{1}{8 v^2} + ... \bigg).
\end{eqnarray}
In the above equations, the terms in the parentheses on the RHS do not pose any problem as they remain finite in this limit. The expressions for the equations of motion for triads become:
\begin{eqnarray}
p_1(t) &=& p_1^0  \exp\left\lbrace\frac{1}{\gamma \lambda} \int_{t_0}^t (\sin(\bar\mu_2 c_2)+\sin(\bar\mu_3 c_3))\cos(\bar\mu_1 c_1) \mathrm{d}t\right\rbrace \nonumber \\
&& \times\exp\left\lbrace \int_{t_0}^t\bigg( \frac{l_o}{\gamma} \sqrt{\frac{p_2 p_3}{p_1^3}}\bigg( 1 - \frac{1}{8 v^2} + ... \bigg)^4\bigg) \cos(\bar\mu_1 c_1) \mathrm{d}t\right\rbrace ,\label{p1-triad-integrated-2} \\
p_2(t) &=& p_1^0  \exp\left\lbrace\frac{1}{\gamma \lambda} \int_{t_0}^t (\sin(\bar\mu_3 c_3)+\sin(\bar\mu_1 c_1))\cos(\bar\mu_2 c_2) \mathrm{d}t\right\rbrace \nonumber \\
&& \times\exp\left\lbrace \int_{t_0}^t\bigg( \frac{l_o}{\gamma} \sqrt{\frac{p_3 p_1}{p_2^3}}\bigg( 1 - \frac{1}{8 v^2} + ... \bigg)^4\bigg) \cos(\bar\mu_2 c_2) \mathrm{d}t\right\rbrace ,\label{p2-triad-integrated-2} ~~~\mathrm{and}~~~\\
p_3(t) &=& p_1^0  \exp\left\lbrace\frac{1}{\gamma \lambda} \int_{t_0}^t (\sin(\bar\mu_1 c_1)+\sin(\bar\mu_2 c_2))\cos(\bar\mu_3 c_3) \mathrm{d}t\right\rbrace \nonumber \\
&& \times\exp\left\lbrace \int_{t_0}^t\bigg( \frac{l_o}{\gamma} \sqrt{\frac{p_1 p_2}{p_3^3}}\bigg( 1 - \frac{1}{8 v^2} + ... \bigg)^4\bigg) \cos(\bar\mu_3 c_3) \mathrm{d}t\right\rbrace .\label{p3-triad-integrated-2}
\end{eqnarray}
Note that in eq. \eqref{p1-triad-integrated-2}, the factor $p_2 p_3/p_1^3$ needs to diverge in order for $p_1$ to either vanish or diverge. In case of $p_2$ and $p_3$ we need the corresponding factors, 
given respectively by  $p_3 p_1/p_2^3$ and $p_1 p_2/p_3^3$, to diverge for them to either vanish or diverge. In case of options A4, A6 and A8, one of the triads is diverging (say $p_1$) while the 
other two remain finite. Since the other two are finite, then the factor $p_2 p_3/p_1^3$ can not be diverging. Hence, $p_1$ cannot be diverging according to eq. \eqref{p1-triad-integrated-2}. 
Similar arguments hold for $p_2$ and $p_3$. Thus, cases A4, A6 and A8 are also inconsistent in this limit. In case A3, all three of the triads diverge. For that to happen according to the above equations of motion we need 
the factors $p_2 p_3/p_1^3$, $p_3 p_1/p_2^3$ and $p_1 p_2/p_3^3$ to diverge simultaneously which is self contradictory. Similarly, we can show that A9 and A10 are also inconsistent with the equations of motion in this limit. 
We are thus left with no viable case in Table I which allows a diverging volume.\\

(iii) {\it{Volume remains finite and non-zero:}} In this case,  options A2, A3, A5, A6, A7 and A9 in Table I are inconsistent. In this case, directly from \eqref{g} we get:
\begin{eqnarray}
g_1(p_1)= \beta^{1/2} (p_2 p_3)^{1/4} ( \sqrt{\vert v+1 \vert} - \sqrt{\vert v-1 \vert}),
\end{eqnarray}
and similarly for $g_2$ and $g_3$, where the terms  in parentheses in RHS remain finite since $v$ is finite. Then, the Hamilton's equations for triads become:
\begin{eqnarray}
p_1(t) &=& p_1^0  \exp\left\lbrace\frac{1}{\gamma \lambda} \int_{t_0}^t (\sin(\bar\mu_2 c_2)+\sin(\bar\mu_3 c_3))\cos(\bar\mu_1 c_1) \mathrm{d}t\right\rbrace \nonumber \\
&& \times\exp\left\lbrace \int_{t_0}^t\bigg( \frac{l_o \beta^2}{\gamma} \sqrt{\frac{(p_2 p_3)^3}{p_1}}( \sqrt{\vert v+1 \vert} - \sqrt{\vert v-1 \vert})^4 \bigg) \cos(\bar\mu_1 c_1) \mathrm{d}t\right\rbrace ,\label{p1-triad-integrated-3} \\
p_2(t) &=& p_2^0  \exp\left\lbrace\frac{1}{\gamma \lambda} \int_{t_0}^t (\sin(\bar\mu_3 c_3)+\sin(\bar\mu_1 c_1))\cos(\bar\mu_2 c_2) \mathrm{d}t\right\rbrace \nonumber \\
&& \times\exp\left\lbrace \int_{t_0}^t\bigg( \frac{l_o \beta^2}{\gamma} \sqrt{\frac{(p_3 p_1)^3}{p_2}}( \sqrt{\vert v+1 \vert} - \sqrt{\vert v-1 \vert})^4 \bigg) \cos(\bar\mu_2 c_2) \mathrm{d}t\right\rbrace ,\label{p2-triad-integrated-3} \\
p_3(t) &=& p_3^0  \exp\left\lbrace\frac{1}{\gamma \lambda} \int_{t_0}^t (\sin(\bar\mu_1 c_1)+\sin(\bar\mu_2 c_2))\cos(\bar\mu_3 c_3) \mathrm{d}t\right\rbrace \nonumber ~~~ \mathrm{and} ~~~,\\
&& \times\exp\left\lbrace \int_{t_0}^t\bigg( \frac{l_o \beta^2}{\gamma} \sqrt{\frac{(p_1 p_2)^3}{p_3}}( \sqrt{\vert v+1 \vert} - \sqrt{\vert v-1 \vert})^4 \bigg) \cos(\bar\mu_3 c_3) \mathrm{d}t\right\rbrace .\label{p3-triad-integrated-3} 
\end{eqnarray}
Let us consider the remaining options in Table I. In options A4 and A8, only one of the triads diverges (say $p_1$). For this to happen, we must have the product $p_2 p_3$ diverging according to eq. \eqref{p1-triad-integrated-3}. But this is inconsistent 
with respect to eqs.(\ref{p2-triad-integrated-3}) and (\ref{p3-triad-integrated-3}). 
Hence, A4 and A8 are ruled out as well. In option A10, one of the triad is vanishing (say $p_1$) while the other two are diverging simultaneously. 
Since the volume is supposed to be finite and non-zero, that means $p_1 p_2 p_3$ is finite and non-zero. Since $p_2$ and $p_3$ are diverging, this means that the products $p_1 p_2$ and $p_1 p_3$ are vanishing. 
But for $p_2$ and $p_3$ to diverge according to above equations, we need the products $p_1 p_2$ and $p_1 p_3$ to diverge. Hence, A10 is also ruled out. 
Therefore, we are left with only option A1, which means all the three triads remain finite and non-zero for all finite time.\\

In summary, we find that out of all possible choices the only physically consistent case is when all the triads remain finite and non-zero for all finite time evolution. Note that this is tied to the inclusion of the inverse 
triad corrections for the `A' quantization, and this result does not arise without the inverse triad corrections. 
We would also like to mention that the same exercise can be repeated for lapse $N=V$ where one obtains identical results.

Let us now consider the energy density. Using the vanishing of the Hamiltonian constraint, it turns out to be 
\begin{eqnarray}		
\rho &=& \frac{1}{8 \pi G\gamma^2 \lambda^2}\bigg[(\sin(\bar\mu_1 c_1)\sin(\bar\mu_2 c_2)+ \mathrm{cyclic}) + l_o \lambda \bigg( \sqrt{\frac{p_1 p_2}{p_3}} g_3^2 \sin(\bar\mu_3 c_3) + \mathrm{cyclic}\bigg)\nonumber \\
&& +\frac{l_o^2 \lambda^2 (1+\gamma^2) (g_1 g_2 g_3)^2}{4 \sqrt{p_1 p_2 p_3}}\bigg(2 p_1^2-p_1^2 p_2^2 g_3^8+ \mathrm{cyclic} \bigg)\bigg] ~.
\end{eqnarray}
Since the triads are non-zero and finite in a finite time evolution, consequently the energy density also remains finite.

The Hubble rates can be evaluated using equations \eqref{hubblerates}. The directional Hubble rate $H_1 $ is given by,
\begin{eqnarray}
H_1 &=& \frac{1}{2 \gamma \lambda} (\sin(\bar\mu_1 c_1 -\bar\mu_2 c_2) + \sin(\bar\mu_1 c_1 -\bar\mu_3 c_3)+\sin(\bar\mu_2 c_2 + \bar\mu_3 c_3) ) \nonumber \\
&& + \frac{l_o}{2\gamma} \bigg(\sqrt{\frac{p_1 p_2}{p_3}}g_3^4 \cos(\bar\mu_3 c_3) + \sqrt{\frac{p_3 p_1}{p_2}}g_2^4\cos(\bar\mu_2 c_2) - \sqrt{\frac{p_2 p_3}{p_1}}g_1^4\cos(\bar\mu_1 c_1) \bigg) ~.
 \end{eqnarray}
We can obtain $ H_2 $ and $ H_3 $  by cyclic permutations of the above equation. We note that the Hubble rates are also bounded and finite as the triads remain bounded and finite for all finite time evolution. 
This implies that the expansion and shear scalars, given by equations \eqref{expansionscalar} and \eqref{shearscalar}, will also be bounded and finite by virtue of directional Hubble rates being finite for all finite time evolution.

Hence we find that in contrast to previous subsection, when we include the inverse triad corrections in the `A' quantization, then the energy density, the Hubble rates, and expansion and shear scalars all remain finite 
for all finite time evolution. However, we still need to consider the curvature invariants and the time derivative of the expansion scalar to analyze their divergence properties. 
We will consider them in Sec. V and show that they can still diverge. But let us first discuss the `K' quantization in the next section.

\section{Effective loop quantum cosmological dynamics: ~~~~~~~~~~~~~~~~~~~~~~~~~~~~~~~~~~~~~~~~~~`K' Quantization}

In this section we consider the effective dynamics of the `K'  quantization. It differs from the `A' quantization in the sense that instead of the Ashtekar-Barbero connection, 
the extrinsic curvature is considered as the momentum variable to the triad to construct holonomies to obtain the field strength. This quantization was first considered for the  Bianchi-IX spacetime 
in Ref. \cite{pswe}, and, the effective Hamiltonian obtained for $ N=1 $ is as follows:
\begin{eqnarray}
\mathcal{H} &=& -\frac{\sqrt{p_1 p_2 p_3}}{8 \pi G\gamma^2}\bigg[\frac{1}{\lambda^2}(\sin(\bar\mu_1 \gamma k_1)\sin(\bar\mu_2 \gamma k_2)+\sin(\bar\mu_2 \gamma k_2)\sin(\bar\mu_3 \gamma k_3)+\sin(\bar\mu_3 \gamma k_3)\sin(\bar\mu_1 \gamma k_1))\nonumber \\
&& +\frac{l_o^2}{4p_1 p_2 p_3}\bigg(2 p_1^2+2p_2^2+2p_3^2-\frac{p_1^2 p_2^2}{p_3^2}-\frac{p_2^2 p_3^2}{p_1^2}-\frac{p_3^2 p_1^2}{p_2^2} \bigg)\bigg]+\rho\sqrt{p_1 p_2 p_3} ~. \label{Hamiltonian}
\end{eqnarray}
Using eq. \eqref{mubar}, the resulting Hamilton's equations for $p_1$ and $k_1$ are:
\begin{eqnarray}
\dot p_1 &=& -8\pi G \frac{\partial\mathcal{H}}{\partial k_1} = \frac{p_1}{\gamma \lambda} (\sin(\bar\mu_2 \gamma k_2)+\sin(\bar\mu_3 \gamma k_3))\cos(\bar\mu_1 \gamma k_1), \label{p1dot} \\
\dot k_1 &=& 8\pi G \frac{\partial\mathcal{H}}{\partial p_1} \nonumber \\
&=& -\frac{\lambda}{2\gamma^2}\frac{1}{\bar\mu_1}\bigg[\frac{1}{\lambda^2}(\sin(\bar\mu_1 \gamma k_1)\sin(\bar\mu_2 \gamma k_2)+\sin(\bar\mu_2 \gamma k_2)\sin(\bar\mu_3 \gamma k_3)+\sin(\bar\mu_3 \gamma k_3)\sin(\bar\mu_1 \gamma k_1))\nonumber \\
&& +\frac{l_o^2}{4p_1 p_2 p_3}\bigg(6 p_1^2-2p_2^2-2p_3^2-3\frac{p_1^2 p_2^2}{p_3^2}+5\frac{p_2^2 p_3^2}{p_1^2}-3\frac{p_3^2 p_1^2}{p_2^2} \bigg)- 8\pi G \gamma^2 (\rho + 2p_1 \frac{\partial \rho}{\partial p_1}) \nonumber \\
&& +\frac{\gamma}{\lambda^2}\bigg(k_1 \bar\mu_1 \cos(\bar\mu_1 \gamma k_1)(\sin(\bar\mu_2 \gamma k_2)+\sin(\bar\mu_3 \gamma k_3)) \nonumber \\
&& -k_2 \bar\mu_2 \cos(\bar\mu_2 \gamma k_2)(\sin(\bar\mu_1 \gamma k_1)+\sin(\bar\mu_3 \gamma k_3)) -k_3 \bar\mu_3 \cos(\bar\mu_3 \gamma k_3)(\sin(\bar\mu_2 \gamma k_2)+\sin(\bar\mu_1 \gamma k_1)\bigg)\bigg].\nonumber \\
\end{eqnarray}
Equations for other phase variables can be found similarly, or obtained from above by  permutations.

Let $t_0 $ be some time in the present at which $p_1$ has some given finite values $p_1^0$. Then from \eqref{p1dot} we get,
\begin{equation}
\int_{p_1^0}^{p_1(t)}\frac{\mathrm{d}p_1}{p_1}= \int_{t_0}^t \frac{1}{\gamma \lambda} (\sin(\bar\mu_2 \gamma k_2)+\sin(\bar\mu_3 \gamma k_3))\cos(\bar\mu_1 \gamma k_1) \mathrm{d}t .
\end{equation}
Upon formal integration, we obtain
\begin{equation}
p_1(t) = p_1^0  \exp\left\lbrace\frac{1}{\gamma \lambda} \int_{t_0}^t\bigg((\sin(\bar\mu_2 \gamma k_2)+\sin(\bar\mu_3 \gamma k_3))\cos(\bar\mu_1 \gamma k_1)\bigg) \mathrm{d}t\right\rbrace .
\end{equation}
\\
Since $ | (\sin(\bar\mu_2 \gamma k_2)+\sin(\bar\mu_3 \gamma k_3))\cos(\bar\mu_1 \gamma k_1)|\leq 2$, the integration (inside the exponential) over a finite time is finite. Hence, for any finite time past or future evolution, we have: 
\begin{equation}
0<p_1(t)<\infty .
\end{equation}
Similarly we can obtain bounds for $p_2$ and $p_3$. Hence, we conclude that
\begin{equation}
0<p_i(t)<\infty \quad \text{and} \quad 0<\frac{1}{p_i(t)}<\infty . \label{p1bound}
\end{equation}
Thus, in contrast to the `A' quantization, the triads in the effective dynamics of `K' quantization never become zero or infinite in a finite time evolution even in absence of the inverse triad modifications.

The expression of energy density is obtained from the vanishing of the Hamiltonian constraint \eqref{Hamiltonian}, yielding 
\begin{eqnarray}
\rho &=& \frac{1}{8 \pi G\gamma^2}\bigg[\frac{1}{\lambda^2}(\sin(\bar\mu_1 \gamma k_1)\sin(\bar\mu_2 \gamma k_2)+\sin(\bar\mu_2 \gamma k_2)\sin(\bar\mu_3 \gamma k_3)+\sin(\bar\mu_3 \gamma k_3)\sin(\bar\mu_1 \gamma k_1))\nonumber \\
&& +\frac{l_o^2}{4p_1 p_2 p_3}\bigg(2 p_1^2+2p_2^2+2p_3^2-\frac{p_1^2 p_2^2}{p_3^2}-\frac{p_2^2 p_3^2}{p_1^2}-\frac{p_3^2 p_1^2}{p_2^2} \bigg)\bigg].
\end{eqnarray}
We find that the energy density is dynamically bounded as a consequence of equation \eqref{p1bound}.

Let us consider the directional Hubble rates. Using equations \eqref{hubblerates}, $ H_1 $ is given by,
\begin{equation}
H_1 = \frac{1}{2 \gamma \lambda} (\sin(\bar\mu_1 \gamma k_1 -\bar\mu_2 \gamma k_2) + \sin(\bar\mu_1 \gamma k_1 -\bar\mu_3 \gamma k_3)+\sin(\bar\mu_2 \gamma k_2 + \bar\mu_3 \gamma k_3) ).
\end{equation}
Thus, we find that $H_1$ is universally bounded. The expressions for $ H_2 $ and $ H_3 $ obtained by cyclic permutations of the above equation yield the same bound. 
As a result, the expansion and shear scalars as given in equations \eqref{expansionscalar} and \eqref{shearscalar} are also universally bounded. 
However, we will discover in the next section that the curvature invariants and the time derivative of the expansion scalar may not necessarily be bounded.

\section{Possible divergences in curvature invariants}

In previous sections we showed that the energy density, expansion and shear scalars, volume and Hubble rates remain non-zero and finite in the effective dynamics of LQC in both the `A' quantization with inverse triad corrections, 
and the `K' quantization. In this section we investigate the potential divergences that might be present in the curvature invariants, and in the time derivative of the expansion scalar $ \dot \theta $ which 
is important to understand the focusing of geodesics. We will show that the curvature invariants and $ \dot \theta $ depend on partial derivatives of the energy density with respect to the triad variables, 
which depend on the matter content. Since we are keeping the matter content arbitrary, these derivatives may diverge for some choices of matter. And if they do, they can cause the curvature invariants to diverge. 
As we discussed in Sec. III and IV, it can be shown that the ratios $\frac{\dot p_i}{p_i}$ remain bounded and that the triads $p_i$ remain positive definite and finite for all finite time 
evolution for both the `A' quantization with inverse triad corrections, and for the `K' quantization. But, we also need to consider the divergence properties of the second derivatives, $\ddot p_i$, as the 
curvature invariants and $ \dot \theta $ depend on them. It is clear from the expressions for $\dot p_i$ as given in previous sections that the second derivatives $\ddot p_i$ will contain first time 
derivatives of the connection variables, $\dot c_i$ or $\dot k_i$. These can be obtained from Hamilton's equations, e.g. :
\begin{equation}
\dot c_i= 8\pi G \gamma \frac{\partial\mathcal{H}}{\partial p_i}.
\end{equation}
We see that $\dot c_i$ will have terms of type $\frac{\partial \rho}{\partial p_i}$. Hence, $\ddot p_i$ and eventually the curvature invariants depend on the partial derivatives of the energy density with respect to the triad 
variables $\frac{\partial \rho}{\partial p_i}$. If $\frac{\partial \rho}{\partial p_i}$ diverge at finite values of volume, energy density, and expansion and shear scalar for some specific choice of matter, 
then the curvature invariants will diverge.

As an illustration, let us consider the `K' quantization discussed in Sec. IV. The second derivative of $p_1$ is given by,
\begin{eqnarray}
\frac{\ddot p_1}{p_1} &=& \bigg(\frac{\dot p_1}{p_1}\bigg)^2 + \frac{\cos(\bar\mu_1 \gamma k_1)}{\lambda}\bigg(\cos(\bar\mu_2 \gamma k_2)(\dot k_2 \bar\mu_2 + k_2 {\dot{\bar{\mu}}_2}) +\cos(\bar\mu_3 \gamma k_3)(\dot k_3 \bar\mu_3 + k_3 {\dot{\bar{\mu}}_3}\bigg) \nonumber \\
&& - \frac{\sin(\bar\mu_1 \gamma k_1)}{\lambda} (\sin(\bar\mu_2 \gamma k_2)+\sin(\bar\mu_3 \gamma k_3))(\dot k_1 \bar\mu_1 + k_1 {\dot{\bar{\mu}}_1}). \label{p1ddot}
\end{eqnarray}
We find that the terms which may lead to divergence are of the form $(\dot k_1 \bar\mu_1 + k_1 {\dot{\bar{\mu}}_1})$, as all other terms and factors are well behaved for any finite time evolution from our conclusions in Sec. IIIB and Sec. IV. 
Let us compute the term $(\dot k_1 \bar\mu_1 + k_1 {\dot{\bar{\mu}}_1})$ to analyze it in detail. It turns out to be:
\begin{eqnarray}
\dot k_1 \bar\mu_1 + k_1 \dot \bar\mu_1 &=& -\frac{\lambda}{2\gamma^2}\bigg[\frac{1}{\lambda^2}(\sin(\bar\mu_1 \gamma k_1)\sin(\bar\mu_2 \gamma k_2)+\sin(\bar\mu_2 \gamma k_2)\sin(\bar\mu_3 \gamma k_3)+\sin(\bar\mu_3 \gamma k_3)\sin(\bar\mu_1 \gamma k_1))\nonumber \\
&& +\frac{l_o^2}{4p_1 p_2 p_3}\bigg(6 p_1^2-2p_2^2-2p_3^2-3\frac{p_1^2 p_2^2}{p_3^2}+5\frac{p_2^2 p_3^2}{p_1^2}-3\frac{p_3^2 p_1^2}{p_2^2} \bigg)- 8\pi G \gamma^2 (\rho + 2p_1 \frac{\partial \rho}{\partial p_1}) \nonumber \\
&& +\frac{\gamma}{\lambda^2}\bigg((k_1 \bar\mu_1-k_2 \bar\mu_2) \cos(\bar\mu_2 \gamma k_2)(\sin(\bar\mu_1 \gamma k_1)+\sin(\bar\mu_3 \gamma k_3)) \nonumber \\
&& +(k_1 \bar\mu_1-k_3 \bar\mu_3) \cos(\bar\mu_3 \gamma k_3)(\sin(\bar\mu_2 \gamma k_2)+\sin(\bar\mu_1 \gamma k_1)\bigg)\bigg]. \label{QOI1}
\end{eqnarray}
We see from the above equation that the divergence can arise only from $\frac{\partial \rho}{\partial p_1}$ and quantities of type $(k_1 \bar\mu_1-k_2 \bar\mu_2)$, 
since all the other quantities are either made of bounded sine and cosine functions or made of the triad variables which will be finite for finite time evolution as shown in Sec. IIIB and Sec. IV. We further note that,
\be
k_1 \bar\mu_1-k_2 \bar\mu_2=\frac{\lambda}{V} (k_1 p_1- k_2 p_2)
\ee
and that,
\begin{eqnarray}
\frac{\mathrm{d}}{\mathrm{d} t} (k_1 p_1- k_2 p_2) &=& \frac{l_o^2}{\gamma^2 V} \bigg(p_2^2 - p_1^2 + \frac{p_3^2 p_1^2}{p_2^2} - \frac{p_2^2 p_3^2}{p_1^2} \bigg)) + 8 \pi G V \bigg(p_1 \frac{\partial \rho}{\partial p_1} - p_2 \frac{\partial \rho}{\partial p_2} \bigg). \label{QOI2}
\end{eqnarray}
In the above equation, the first term is bounded. Hence, the quantity $(k_1 p_1- k_2 p_2)$ (and consequently $k_1 \bar\mu_1-k_2 \bar\mu_2$), which is the integration of the right hand side of \eqref{QOI2}, 
may only diverge if the partial derivatives of the energy density with respect to triads diverges.

Therefore, we conclude from our discussion of equations \eqref{p1ddot},\eqref{QOI1} and \eqref{QOI2}, that the quantities of the form $(\dot k_i \bar\mu_i - k_j {\dot{\bar{\mu}}_j})$, and consequently the second 
derivatives $\frac{\ddot p_i}{p_i}$, and hence the curvature invariants and $ \dot \theta $ will only diverge when the partial derivatives of the energy density with respect to triad variables diverge at a finite value of energy density. 
Such conditions require highly exotic equations of state of matter. Also note that in case of matter with vanishing anisotropic stress, the quantities $\frac{\partial \rho}{\partial p_i}$ are proportional to the pressure. 
So we can call these divergences in the curvature invariants and $ \dot \theta $ as ``pressure divergences".

Above we focused on the `K' quantization. However, one reaches the same conclusion by repeating this exercise for the second derivatives of triad variables in the `A' quantization with inverse triad corrections. To summarize, we note 
that the curvature invariants or the time derivative of the expansion scalar may not be bounded in effective dynamics. Some quantities of interest may diverge both in `A' and `K' quantizations. 
However, as we will see in the next two sections, that geodesics do not break down at such events if such divergences occur at a finite time, and that these divergences do not amount to strong singularities. Therefore, such potential curvature 
divergences will turn out to be harmless events.

\section{Geodesic Completeness}
Given a spacetime metric, the geodesic equations are given by:
\be
(x^i)'' = \Gamma^{i}_{jk}(x^j)'(x^k)' . \label{gdesics1}
\ee
where prime denotes derivative with respect to the affine parameter. These equations give us the accelerations along the geodesic curves in terms of the Christoffel symbols and the velocity four vector along the geodesic. 
We can integrate them to find out the expression for the velocities as follows, 
\be
(x^i)' = \int \mathrm{d}\tau \bigg(\Gamma^{i}_{jk}(x^j)'(x^k)'\bigg). \label{gdesics2}
\ee
Let us note that for comoving observers $t$ itself is the affine parameter. The coefficients $\Gamma^{i}_{jk}$ are obtained from first derivatives of the metric, hence cannot depend on the second time derivatives of the triads. 
They only depend on the triads and their first time derivatives and the angular variables in the Bianchi-IX metric. Using the relations between triads and scale factors, we can express them in terms of the scale factors and their first time derivatives. 
Using our results in previous sections, the  triads and their first time derivatives (and hence the scale factors and their first time derivatives) remain finite for all finite values of time $t$, which is the affine parameter. 
Hence, the coefficients $\Gamma^{i}_{jk}$ as functions of the affine parameter will be finite for any finite value of the affine parameter for both the `A' quantization with inverse triad modifications, and `K' quantization. 

In the classical dynamics of Bianchi-IX spacetime, the cosmological singularity is encountered due to vanishing of one or more scale factors while the rest are either diverging or remain finite. 
There is an associated breakdown of geodesics as some of the terms in the expressions for the accelerations and velocities along the geodesics contain powers of the scale factors or their first derivatives as discussed above. 
To see this, we consider the Bianchi-IX metric in presence of local rotational symmetry which simplifies the Christoffel symbols considerably. The metric takes the following form:
\be
\mathrm{d}s^2= - \mathrm{d}t^2 + a_1^2 \mathrm{d}x^2 + a_2^2 \mathrm{d}y^2 + (a_2^2 \sin^2 y + a_1^2 \cos^2 y) \mathrm{d}z^2 - 2 a_1^2 (\cos y) \mathrm{d}x \mathrm{d}z .
\ee
For this metric, the geodesic equations can be simplified to the following form in this case:
\begin{eqnarray}
x'' &=& -2\frac{a'_1}{a_1^3} - \frac{C}{a_2^2} (\cot{y}) y' - 2 z' \cos{y} \bigg(\frac{a'_2}{a_2}+ y' \cot{y} \bigg) - y'z' \sin{y}, \\
y'' &=& -2\frac{a'_2}{a_2} y' + z'^2 \cos{y} \sin{y} + \frac{C}{a_2^2} z' \sin{y}, \\
z'' &=& -2\frac{a'_2}{a_2} z' - 2 y'z' \cot{y}  - \frac{C}{a_2^2} y' \csc{y} ,
\end{eqnarray}
where $C$ is a constant. We see that these geodesics have the scale factors and their first derivatives as coefficients, and will break down at the classical cosmological singularity precisely because one or more of these coefficients diverge.

The geodesic evolution in effective dynamics of `A' quantization with inverse triad effects, and 'K' quantization differs significantly from the classical theory in the evolution of the scale factors. 
In the effective dynamics in above cases, as we discussed in the previous paragraph, these scale factors and their first derivatives remain finite for any finite value of the affine parameter. 
Effective dynamics modifies the spacetime in such a way that as we approach the classical cosmological singularity, the scale factors, instead of vanishing or diverging, reach some finite limiting value and then bounce back. 
The singularity is replaced by a bounce such that the scale factors are well behaved. As a result, the classical cosmological singularity is absent in the effective spacetime.


We found in the previous section that some of the curvature invariants may diverge if the derivatives of the energy density with respect to triad variables diverge. 
However if such divergence in curvature invariants happens in a finite time evolution, then from the results of this section we know that the geodesics will not break down at such events as the Christoffel symbols remain 
finite for any finite time evolution, hence the velocities and accelerations along the geodesics will be well defined. This means that such divergences in curvature invariants, if they happen in a finite time, do not amount to the breakdown of geodesics. 

In summary, for the case of `A' quantization with inverse triad effects included, and the `K' quantization, geodesics will be well-behaved in effective spacetime at all those events where they could break down classically due 
to divergences in scale factors and their first derivatives in finite time evolution. In this sense we can say that the Bianchi-IX spacetime is geodesically complete in the effective dynamics of LQC for the `A' quantization with inverse triad effects and the `K' quantization.
In the case of `A' quantization without the inclusion of inverse triad effects, scale factors can vanish and diverge in finite time and geodesic evolution will not be complete.

\section{Lack of Strong Singularities in effective dynamics}

In the previous section we have shown that geodesics will go past the events where divergences occur in curvature invariants in finite time evolution for the `A' quantization with inverse triad modifications and for the `K' quantization.
We now show that these divergences do not amount to a strong curvature singularity. A strong curvature singularity is defined by analyzing the fate of an extended object as it falls into the singularity.
Due to the curvature, adjacent radial geodesics get more and more focused as we approach the singularity. This means that any extended object falling into the singularity, which can be considered as a system of particles 
following adjacent radial geodesics, will get squeezed as it falls into the singularity. If the tidal forces are strong enough that the any object falling into the singularity is crushed down to zero volume regardless of the properties 
of the falling object, then it is referred to as a strong singularity \cite{ellis1977singular,tipler1977singularities,ck-1985conditions}. A necessary condition for a singularity to be a strong singularity was derived by 
Tipler \cite{tipler1977singularities}. Clarke and Krolak \cite{ck-1985conditions} generalized Tipler's condition and arrived at the following necessary condition for a singularity to be a strong singularity:

If a singularity is a strong curvature singularity, then, for a timelike (or null) geodesic running into the singularity, the integral
	\begin{equation}
	K^{i}_{j} =\int_0^{\tau} \mathrm{d}\tau' |R^{i}_{4j4} (\tau')| \label{krolak}
	\end{equation}
	does not converge as $\tau \rightarrow \tau_{o} $, where $\tau_{o} $ is the position of singularity. Otherwise the singularity is termed as a weak singularity.
	
The integral \eqref{krolak} involves one of the Riemann curvature tensor components. Let us investigate the behavior of Riemann curvature tensor to gain insight into this. The Riemann curvature tensor is obtained from second 
derivatives of the metric. The metric is a function of time through the triad variables, and is a function of other three angular variables in terms of their sines or cosines in the numerator. 
When we compute all possible second derivatives of the metric, all the derivatives with respect to the angle variables again result into sines and cosines of those variables in the numerator. 
Hence, the curvature tensors will be made of only the following three types of terms:\\

T1) Terms of type $f(p_1,p_2,p_3)$, obtained when both the derivatives were with respect to the angle variables. These terms may also contain sines and cosines of the angle variables in their numerators. \\

T2) Terms of type $\bigg(\frac{\dot p_1}{p_1}\bigg)^m \bigg(\frac{\dot p_2}{p_2}\bigg)^n \bigg(\frac{\dot p_3}{p_3}\bigg)^q f(p_1,p_2,p_3)$ where $m,n,q$ are positive integers. 
These terms are obtained when either one or both of the derivatives is with respect to time.\\

T3) Terms of type $\frac{\ddot p_i}{p_i}f(p_1,p_2,p_3)$, where $i$ can be $1,2$ or $3$, obtained when both the derivatives are with respect to time.\\

The components of the curvature tensors are made of sums or differences of these three types of terms. Let us look at the behavior of each type of term under the integral \eqref{krolak}.

Since we have shown that all the triad variables are bounded for any finite evolution, and we see from \eqref{p1dot} that $\frac{\dot p_i}{p_i}$ is a bounded function, terms of type (T1) and (T2) are finite for finite time evolution. 
Since they themselves are bounded, they give a finite quantity when integrated over a finite interval. On the other hand, terms of type T3 contain the quantities $\frac{\ddot p_i}{p_i}$ which may lead to divergences. 
However, upon integration in \eqref{krolak}, these second derivatives of the triads are removed as follows:
\be
\int_0^{\tau_o} \frac{\ddot p_i}{p_i}f(p_1,p_2,p_3)\mathrm{d}\tau  = \frac{\dot p_i}{p_i}f(p_1,p_2,p_3) \bigg\vert_0^{\tau_o} - \int_0^{\tau_o} \dot p_i \bigg(\frac{d}{\mathrm{d}\tau}f_1(p_1,p_2,p_3)\bigg)\mathrm{d}\tau .
\ee
We see that both the terms on the RHS are finite, hence the terms of type (T3) also do not lead to divergence upon integration over a finite interval. 

Since these terms (T1), (T2) and (T3) individually do not lead to divergence upon integration over a finite interval, we can conclude that the integral \eqref{krolak} will also remain finite 
because the integral of the absolute value of a sum of terms is less than equal to the sum of integrals of individual terms:
\begin{equation}
\int_a^b |T_1 + T_2 + ... + T_n| \mathrm{d}t \leq \int_a^b |T_1|\mathrm{d}t + \int_a^b |T_2|\mathrm{d}t + ..... + \int_a^b |T_n| \mathrm{d}t .
\end{equation}

We have hence shown that the necessary conditions for existence of strong singularities are not satisfied in the Bianchi-IX model in effective dynamics of LQC for the `A' quantization with inverse triad corrections and for the `K' quantization. 
This means that any singularities, if present, are of weak curvature type. We have already shown in the previous section that geodesics can be extended beyond such events. Hence, these singularities turn out to be harmless. 

\section{Conclusion}

In the last decade, there have been many developments on the resolution of classical singularities due to quantum gravity effects originating from loop quantum gravity.
 There has been significant progress in showing the generic resolution of singularities in various symmetry reduced spacetimes, starting from isotropic and homogeneous flat spacetime \cite{ps09}, and then going to 
 increasingly more anisotropic spacetimes to include Bianchi-I spacetime \cite{ps11}, Kantowski-Sachs spacetime \cite{ks-strong} and Bianchi-II spacetime \cite{Saini2017}. 
 In continuation of this line of work, the natural step forward was to consider the Bianchi-IX spacetime, which is the subject matter of this manuscript. In particular, our goal in this manuscript was to generalize the result 
 on resolution of strong singularities to the Bianchi-IX spacetime in the effective dynamics of LQC. The studies of BKL conjecture and Mixmaster universe suggest that Bianchi-IX spacetime describes the dynamics of a generic collapse 
 in the close vicinity of the singularity \cite{bkl1971, Berger1998}. For this reason, Bianchi-IX spacetime is an important model to consider for singularity resolution. 
 
 We studied the effective dynamics of Bianchi-IX spacetime in LQC under two different quantizations - the connection operator based `A' quantization, and the extrinsic curvature based `K'  quantization. 
 Both these quantizations were also considered in the Bianchi-II case \cite{Saini2017}, and it was found that the `K'  quantization provided a much cleaner path to singularity resolution than the `A' quantization. At the effective spacetime level, 
 there are two different effects coming from quantum gravity that modify the Hamiltonian constraint - the holonomy corrections and the inverse triad corrections. In the cases of the `A' quantization, in the cases 
 of isotropic and homogeneous spacetimes, Bianchi-I spacetime and Kantowski-Sachs spacetime with arbitrary minimally coupled matter, the holonomy corrections were enough to resolve the singularities and one could do 
 without including inverse triad corrections. But, in Bianchi-II spacetime it was also found for the first time that inverse triad corrections cannot always be ignored in the `A' quantization. 
 It was seen that singularity resolution was not possible to obtain without including both holonomy and inverse triad corrections in the effective Hamiltonian constraint in the `A' quantization. 
 However, there was one important caveat to this in the Bianchi-II case. One could ignore the inverse triad corrections and still obtain singularity resolution if weak energy conditions are imposed on the matter content, 
 a reasonable condition on the matter content also used in the singularity theorems in classical GR. However, we found in this manuscript that the generalization to the `A' quantization of Bianchi-IX spacetime removes that 
 caveat and brings the role of inverse triad corrections 
 in resolving the singularities to the center stage. We saw in Sec. IIIA that the `A' quantization without inverse triad corrections does not prevent the triad variables (and hence volume) from vanishing in a finite time evolution 
 and leads to an unbounded energy density, Hubble rates and expansion and shear scalars. Further, unlike the Bianchi-II case, even imposing weak energy conditions is not sufficient to avoid these scenarios in the `A' quantization if inverse triad corrections are ignored. 
 In Sec. IIIB we included the inverse triad corrections and considered all the possible scenarios that might lead to singularities. We showed that all the triads remain finite and non-zero for all finite time evolution by 
 including the inverse triad corrections which leads to expansion and shear scalars and Hubble rates being finite for any finite time evolution. The energy density also remains finite consequently for all finite time evolution.

In contrast to the above results, the `K' quantization proves to be a superior alternative to the `A' quantization. 
In the Bianchi-II case, it was seen that geodesic completeness and resolution of strong singularities in the `K' quantization could be obtained in a straight forward manner without including inverse triad corrections 
or imposing energy conditions on the matter content. The advantages of the `K'  quantization had already been discussed in Ref. \cite{pswe}, where it was shown to lead to generically bounded expressions for the expansion and shear scalars. 
In the present manuscript we take these results further to the case of Bianchi-IX spacetime. In Sec. IV we contrast the behavior of triad variables in the `K'  quantization as compared to the `A' quantization 
and find that they remain finite and non-zero for all finite time evolution, which implies that the energy density is also finite for all finite time evolution, and expansion and shear scalars and Hubble rates are generically bounded for all time.

Both, the `A' quantization with inverse triad corrections and the `K'  quantization, do not prevent the curvature invariants from diverging as was found in Sec. V. These divergences arise from the blow up of  derivatives of energy density at a finite energy density 
and require highly exotic equations of state.
However,  we further showed in Sec VI that geodesics are well behaved when such divergences occur in the curvature invariants in finite time evolution. 
We also noted that geodesics are also well-behaved at the classical singularity which in LQC is replaced by multiple bounces of scale factors. Finally, in Sec. VII we showed that these divergences in curvature invariants 
do not amount to strong curvature singularities. Such curvature divergent events turn out to be weak curvature singularities beyond which geodesics can be extended.

Our analysis shows that if we consider the `A' quantization, inverse triad corrections are expected to play a crucial role in a potential `non-singularity theorem' in LQC. And, it is also clear that energy conditions 
are unlikely to play any significant role in such a theorem, even though the singularity theorems of classical GR rely on these energy conditions. That means any such theorem would hold for an arbitrary matter content without any energy conditions, i.e. quantum geometric effects of gravity would by themselves avoid singularities
regardless of the matter content. It must be noted that the `K'  quantization provides an alternative to the `A' quantization where inverse triad corrections can be ignored so far. The quantization provides a simpler route to singularity resolution in LQC. 
Following this path of quantization and taking into account results from different anisotropic spacetimes,  there is no indication that energy conditions on the matter content  would need to be imposed in a potential `non-singularity theorem.'

\begin{acknowledgments}
This work is supported by NSF grants PHYS1404240 and PHYS1454832.
\end{acknowledgments}


\end{document}